\documentclass[structabstract]{aa}  
\usepackage{graphicx}
\usepackage{txfonts}
\usepackage{amssymb}
\usepackage{natbib}
\bibpunct{(}{)}{;}{a}{}{,}
\usepackage{ulem}
\usepackage{color} 

\newcommand{\G}{$\Gamma$}
\newcommand{\NH}{$N_{\rm H}$}
\newcommand{\Mjup}{$\mathrm{M_{J}}$}

\newcommand{\mMsun}{\mathrm{M_{\sun}}}
\newcommand{\mMjup}{\mathrm{M_{J}}}
\newcommand{\mbh}{$M_{\rm BH}$}
\newcommand{\mmbh}{M_{\rm BH}}
\newcommand{\mstar}{$M_{\star}$}
\newcommand{\rstar}{$R_{\star}$}
\newcommand{\cnts}{count s$^{-1}$}
\newcommand{\ergs}{erg s$^{-1}$}
\newcommand{\ergscm}{erg cm$^{-2}$ s$^{-1}$}
\newcommand{\kmsMpc}{km s$^{-1}$ Mpc$^{-1}$}
\newcommand{\soft}{$\mathrm{Soft}$}
\newcommand{\hard}{$\mathrm{Hard}$}
\newcommand{\xmm}{\textit{XMM-Newton}}
\newcommand{\integral}{\textit{INTEGRAL}}
\newcommand{\swift}{\textit{Swift}}
\newcommand{\maxi}{\textit{MAXI}}
\newcommand{\einst}{\textit{EINSTEIN}}
\newcommand{\isgri}{IBIS/ISGRI}
\newcommand{\igrj}{IGR~J12580+0134}
\newcommand{\ngc}{NGC~4845}

\begin{document}

\title{Tidal disruption of a super-Jupiter by a massive black hole}

\author{M. Niko\l ajuk
            \inst{1,2}
            \and
            R. Walter\inst{1}
}

\institute{ISDC Data Centre for Astrophysics, Observatoire de Gen\`eve, Universit\'e de Gen\`eve,
           Chemin d'Ecogia 16, CH-1290 Versoix, Switzerland \\
    \email{mrk@alpha.uwb.edu.pl} \\
    \email{Roland.Walter@unige.ch}
  \and
          Faculty of Physics, University of Bialystok, Lipowa 41, 
          15-424 Bia\l ystok, Poland \\
}

\date{Received October 30, 2012; accepted February 13, 2013}

 
  \abstract
   {}
   {A strong, hard X-ray flare was discovered (\igrj) by \integral\ in
     2011, and is associated to \ngc, a~Seyfert~2 galaxy
     never detected at high-energy previously.  To understand
     what happened we observed this event in the X-ray band on several
     occasions.  }
   {Follow-up observations with \xmm, \swift, and \maxi\ are presented
     together with the \integral\ data.  Long and short term
     variability are analysed and the event wide band
       spectral shape modeled. }
   {The spectrum of the source can be described with an absorbed
     ($N_{\rm H} \sim 7 \times 10^{22}$ cm$^{-2}$) power law ($\Gamma
     \simeq 2.2$), characteristic of an accreting source, plus a soft
     X-ray excess, likely to be of diffuse nature. The hard X-ray flux
     increased to maximum in a few weeks and decreased over a year,
     with the evolution expected for a tidal disruption event. The
     fast variations observed near the flare maximum allowed us to
     estimate the mass of the central black hole in \ngc\ as $\sim
     3\times 10^5 \mMsun$. The observed flare corresponds to the
     disruption of about 10\% of an object with a mass of 14-30
     Jupiter.  The hard X-ray emission should come from a corona
     forming around the accretion flow close to the black hole. This
     is the first tidal event where such a corona has been observed.  }
   {}


   \maketitle
%
\section{Introduction}

NGC~4845 is a nearby, high surface brightness spiral galaxy with
morphological type {\it SA(s)ab~sp} classified as a Seyfert~2
\citep{VV2006}.  The distance to \ngc\ was determined as $D=15.6$ Mpc
based on the Tully-Fisher relation \citep{TF88}. \citet{Corsini99}
obtained $D=13.1$ Mpc by including the influence of a dark matter to
the systematic velocity.  We adopt in this paper $D=14.5$ Mpc
($H_0=75$ \kmsMpc) after \citet{Shapley2001} who performed a thorough
literature search to determine the most reliable distance of this
galaxy.

The galactic disk inclination is 76\degr\ \citep{Pizzella2005}.
Coordinates of the galaxy (i.e. $\alpha_{\rm J2000}=12^{\rm h} 58^{\rm
  m} 01\fs19$ and $\delta_{\rm J2000}=+01\degr 34\arcmin 33\farcs02$,
\citealt{VV2010}) set the Galactic neutral hydrogen column density
into the direction of the source to $\mathrm{N_H^{Gal}} = 1.67
\times 10^{20}$ cm$^{-2}$ \citep{DL90,Stark92}.

\ngc\ has been observed, so far, in the optical (e.g. by the Sloan Digital
Sky Survey -- SDSS, the Palomar Optical Sky Survey -- POSS and the
Hubble Space Telescope -- HST), and in the infrared domain (by the Two
Micron All Sky Survey -- 2MASS and the Infrared Astronomical Satellite 
-- IRAS)
\citep{Moshir90,Spinoglio95,Djorgovski98,Sanders2003,Jarrett2003,
Schneider2005}.

The galaxy has also been observed by the Green Bank Telescope and the Very
Large Array in the radio bands 1.4, 4.8, 8.4~GHz
\citep{Condon98,Condon2002}.  The radio emission is dominated by
star-forming regions \citep[see][their Fig. 5d]{Filho2000} and
does not reveal water maser emission \citep{Braatz2003}.
Monitoring in the X-ray domain was carried out using the Imaging
Proportional Counter onboard the \einst\ satellite
\citep{Fabbiano92}.  The authors estimated only an upper limit in the
0.2-4.0 keV energy band of $F_{\rm X} < 2.52\times 10^{-13}$ \ergscm.

\integral\ discovered a new hard X-ray source \igrj\ \citep{ATel2011}
during an observation performed in the period January 2-11, 2011, with
a position (i.e. $\mathrm{RA=194.5212}$ deg $\mathrm{DEC=1.5738}$ deg,
$\pm 2.3$ arcmin, J2000.0) consistent with that of \ngc.  A few
days later, \swift/XRT and \xmm\ observations confirmed the association
with the central regions of the Seyfert~2 galaxy.  \igrj\ was detected
at a peak flux $F_{\rm 2-10 keV} > 5.0\times 10^{-11}$ \ergscm,
corresponding to a variability by a factor $>100$, which is very
unusual for a Seyfert 2 galaxy.

In sect.~\ref{sec:data} we report the analysis of the available
high-energy observations of \igrj\ performed with \xmm, \swift, \maxi,
and \integral.  We analyse these results in sect.~\ref{sec:res} and
discuss them in sect.~\ref{sec:dis}.

%

\section{High-energy observations of \igrj}
\label{sec:data}

We analysed \integral, \maxi, \swift, and \xmm\ observations of \igrj.
A~log of the pointed observations is given in Tables~\ref{tab:log} and
\ref{tab:log2}.  Throughout this section, all uncertainties are given
at 1~sigma confidence level.

\begin{table}
\caption[]{\swift\ and \xmm\ observations of IGR J12580+0134.}
\label{tab:log}
\centering
\begin{tabular}{l c c c} 
\hline
\hline
Instrument & Obs. ID & Start time  & Net exp. \\
 & & (UTC) & (ks) \\
\hline
\swift & 00031911001 & 2011-01-12~23:59:01 & 3.0 \\
\swift & 00031911002 & 2011-01-13~03:26:01 & 2.1 \\
\xmm   &  0658400601 & 2011-01-22~16:23:28 & 14.0 \\
\swift & 00031911003 & 2012-06-29~07:56:59 & 3.3 \\
\hline
\end{tabular}
\end{table}
\begin{table*}
\caption[]{\integral\ observations of IGR J12580+0134.}
\label{tab:log2}
\centering
\begin{tabular}{l c c c c c} 
\hline
\hline
Revolutions & \multicolumn{2}{c}{Start time} & Stop time & EXP \\
                                & (UTC) & (MJD) & (UTC) & (ks) \\
\hline
1004-1006 & 2011-01-02 13:51:14 & 55563.57725 & 2011-01-11 04:47:40 & 100.0 \\
1009      & 2011-01-18 19:09:42 & 55579.79841 & 2011-01-22 01:49:00 & 1.6  \\
1012      & 2011-01-27 10:55:12 & 55588.45500 & 2011-01-27 11:21:46 & 1.0  \\
1052-1055 & 2011-05-28 00:16:37 & 55709.01155 & 2011-06-06 03:15:17 & 44.2 \\
1057-1061 & 2011-06-10 03:14:24 & 55722.13500 & 2011-06-24 03:55:16 & 90.4 \\
1063      & 2011-06-28 12:02:32 & 55740.50176 & 2011-06-30 17:10:01 & 29.0 \\
1067-1068 & 2011-07-10 21:03:15 & 55752.87726 & 2011-07-15 16:05:07 & 45.0 \\
1070      & 2011-07-20 08:31:40 & 55762.35533 & 2011-07-21 16:29:40 & 11.7 \\
\hline
\end{tabular}
\tablefoot{EXP indicates the effective exposure time.}
\end{table*}
\subsection{INTEGRAL}

\integral\ data were analysed using OSA software (ver. 9). We
considered data from \isgri\ in the 17.3-80 keV. The average fluxes
and spectrum were extracted from the mosaic images with
\texttt{mosaic\_spec}.

\integral\ observations are divided into pointings with duration of
$\sim$ 2-3 ks, which are collected during the satellite revolutions. 
We considered in this paper all available (i.e. 555)
pointings from the discovery observation to July 21, 2011, included in
revolutions 1004-1070. 

\begin{figure}
\centering\includegraphics[width=0.5\textwidth]{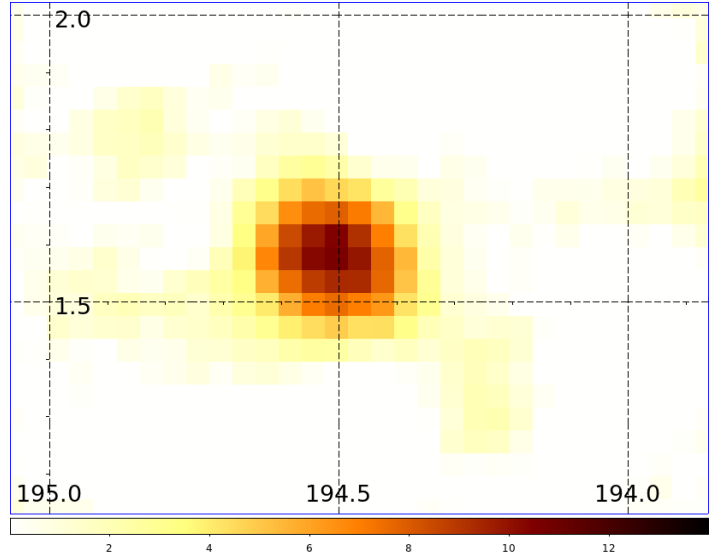}
\caption{The \integral\ IBIS/ISGRI mosaic of \igrj\ 
(17.3-80 keV, significance map) observed 2-11 January 2011.
The dash black grid denotes equatorial J2000.0 coordinates in degrees.
}
\label{fig:integralobs}
\end{figure}

Figure~\ref{fig:integralobs} shows the IBIS/ISGRI
\citep{Lebrun2003,Ubertini2003} mosaic obtained around \igrj\ (17.3-80
keV, significance map) accumulated from January 2 to 11, 2011. During
this period the source significance reached 9.7 sigma in the 20-40 keV
energy band, corresponding to a flux of $F_{\rm 20-40 keV} = (2.5 \pm
0.3) \times 10^{-11}$ \ergscm\ for an exposure time of 400 ks. The
source was also detected by JEM-X \citep{Lund2003} (3-20~keV) for an
effective exposure time of \mbox{13 ks} with a flux of $F_{\rm 3-10
  keV} = (5.0 \pm 1.4) \times 10^{-11}$ \ergscm.
  
We also analysed all publicly available \isgri\ data obtained on
the field from the beginning of the mission up to January 27, 2011
using the HEAVENS interface \citep{Walter2010} with OSA9.
The \integral\ observations performed in previous years (i.e. 
from 2003 to 2010) did not reveal any hard X-ray activity, and we 
obtained a five-sigma upper limit of $5.9 \times 10^{-12}$ \ergscm\ in the 
20-40 keV energy band.  

\begin{figure}
\centering
\includegraphics[width=0.5\textwidth]{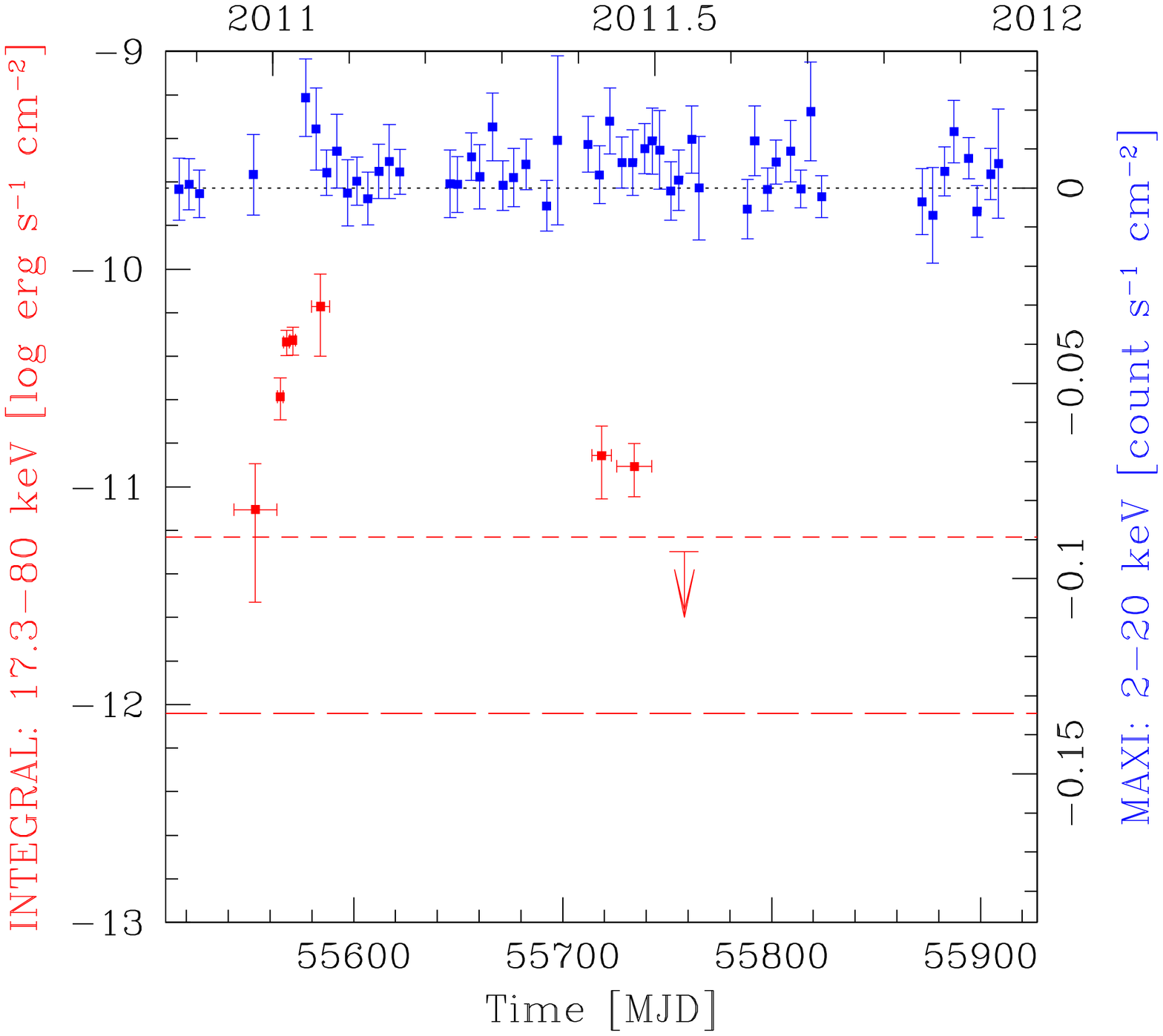}
\caption{The light curve of \igrj\ observed in the 17.3-80 keV energy
  band by \integral/ISGRI (red points) and in the 2-20 keV band
  by \maxi\ (blue points). The arrow shows an upper limit.
  Red short-dash line indicates the $5\sigma$ upper limit of the
  source observed by \isgri\ before 2011. Red long-dash line shows an
  upper limit observed by \einst\ \citep{Fabbiano92} and interpolated
  to 17.3-80 keV. The interpolation is based on the model fitted to
  the \xmm/\integral\ observation.  The interpolated \einst\ flux
  is very likely overestimated because of the harder spectrum observed in
  2011.  }
\label{fig:maxiISGRI}
\end{figure}

The resulting count rates are listed in  Table \ref{tab:lcINTEGRAL} and
plotted in Fig.~\ref{fig:maxiISGRI}.
 
\begin{table}
\caption[]{Intensities of IGR J12580+0134 observed by \integral /ISGRI.}
\label{tab:lcINTEGRAL}
\centering
\begin{tabular}{c c c c c} 
\hline
\hline
START TIME & & END TIME & Count rate  & Signific. \\
(MJD) & & (MJD) & (count s$^{-1}$) & \\
\hline
\multicolumn{5}{l}{{\bf During flare (i.e. Jan 2011)}} \\
55542.63936 &$^a$& 55563.22030 & $0.192 \pm 0.120$ & 1.6 \\ 
55563.57725 &    & 55566.23899 & $0.635 \pm 0.139$ & 4.6 \\
55566.56943 &    & 55569.20899 & $1.132 \pm 0.153$ & 7.4 \\
55569.56013 &    & 55572.19978 & $1.157 \pm 0.167$ & 6.9 \\
55579.79841 &$^a$& 55588.47273 & $1.652 \pm 0.674$ & 2.4 \\
\multicolumn{5}{l}{{\bf During flare -- total}} \\
55563.57725 &    & 55572.19978 & $0.944 \pm 0.087$ & 10.8 \\
\hline
\multicolumn{5}{l}{{\bf After flare (i.e.  Jun-Jul 2011)}} \\
55713.98265 &    & 55723.36030 & $0.341 \pm 0.125$ & 2.7 \\
55725.96026 &    & 55742.71529 & $0.304 \pm 0.083$ & 3.7 \\
55752.87726 &    & 55763.68727 & $0.013 \pm 0.111$ & 0.1 \\
\multicolumn{5}{l}{{\bf After flare -- total}} \\
55713.98265 &    & 55763.68727 & $0.269 \pm 0.056$ & 4.8 \\
\hline
\multicolumn{5}{l}{{\bf Period} $\mathbf{< 2011}$ {\bf (Jan 2003-Dec 2010)} } \\
52644.50424 &$^a$& 55545.30596 & $0.014 \pm 0.026$ & 0.5 \\
\hline
\multicolumn{5}{l}{{\bf Period} $\mathbf{> 2011}$ {\bf (Jan-Jul 2011)} } \\
55563.57724 &    & 55763.68724 & $0.470 \pm 0.047$ & 9.9 \\
\hline
\end{tabular}
\tablefoot{
$^a$ Data obtained through the HEAVENS interface.}
\end{table}

\subsection{MAXI}

\ngc\ has been observed by the Monitor of All-sky X-ray Image (\maxi),
attached to the Japanese Experiment Module onboard the
International Space Station. Among the two X-ray instruments onboard
MAXI, we analysed only the Gas Slit Camera (GSC) data in the present
paper, because of its higher sensitivity \citep{Matsuoka2009,Mihara2011}.

Figure~\ref{fig:maxiISGRI} shows the \maxi\ light curve of \ngc\ in the
2-20 keV band, obtained from the \maxi/GSC on-demand process interface
with the source and background radii of 2 and 3 deg, respectively,
centred on the galaxy.

\subsection{Swift}

The field of view (FOV) around \igrj\ was observed three times with
\swift/XRT on January 12 and 13, 2011 and on June 29, 2012 for a total
exposure of 3.1, 2.1, 3.3 ks (observations ID 00031911001, 00031911002,
and 00031911003, PI Walter), respectively.  We analysed data collected
in photon-counting mode (PC) by using standard procedures
\citep{Burrows2005} and the latest calibration files. The \swift/XRT
`enhanced' position of the source is $\alpha_{\rm J2000} = 194.50492$
deg and $\delta_{\rm J2000}=1.57579$ deg with the error of 2.1 arcsec.
To obtain those values we followed the procedure described in
\citet{Evans2009} where we used the \swift/XRT-\swift/UVOT alignment
and matching UVOT FOV to the USNO-B1 catalogue.  We extracted
light curves and spectra of \igrj\ using the Level 2 event
files\footnote{The online XRT analysis threads
  http://www.swift.ac.uk/XRT.shtml}.  No pile-up problems were found
in either observation.  Therefore, we used circle region with radius of
20 pixels ($\sim$47 arcsec) around the source coordinates. The
background region, which lies in the no visible sources area, was also
chosen as a circle with a radius of 50 pixels. There are a number of
bad columns on the XRT CCD, and the corresponding pixels were not used
to collect the data.  To account for them, we used a corrected
exposure map created with the {\sc xrtexpomap} command and used it to
create the ancillary response files with the command
\texttt{xrtmkarf}. We maximised the signal-to-noise (S/N) by
summing up all the available data obtained in January 2011 (effective
exposure time 5116 s). The final extracted spectrum was rebinned to
have a minimum of 30 counts in each energy bin. Only 12 source photons
were detected in June 2012.

\subsection{XMM-Newton}

\begin{figure}
\centering
\includegraphics[width=0.5\textwidth]{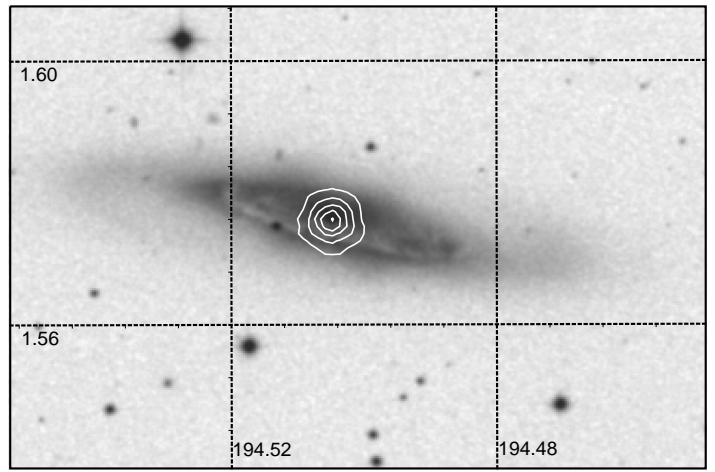}
\caption{The optical finding chart of \ngc\ taken from 
the Space Telescope Science Institute - Digitized Sky Survey (STSci-DSS).
White solid contours show the \xmm\ Epic-pn image of \igrj\ 
(0.8-10 keV, significance map). The black grid denotes 
equatorial J2000.0 coordinates in degrees.
}
\label{fig:xmmobs}
\end{figure}

\xmm\ observed \igrj\ on January~22, 2011 (obs. ID 0658400601,
total exposure time $\sim 21$ ks, PI Walter) and obtained a source position of 
$\alpha_{\rm J2000} =194.50427$ deg and $\delta_{\rm J2000}=1.57585$ deg 
4.1 arcsec (Fig.~\ref{fig:xmmobs}).

During this observation all the EPIC-pn and EPIC-MOS cameras were 
operated in full frame mode. The galaxy was observed with both the EPIC-pn 
and EPIC-MOS; however, in this paper we focus on the data obtained from 
the first detector.

We processed \xmm\ observation data files (ODFs) with pipeline
\textsc{eproc} (Science Analysis System, SAS, v.10.0) in order to
produce a calibrated event list. The event file was filtered to exclude
high background time intervals following the recommendations of the
SAS analysis thread\footnote{How to extract PN spectra see also
http://xmm.esa.int/sas/current/documentation/threads/PN\_spectrum\_thread.shtml}. 
We extracted the light curve in the 10-12 keV energy band using all
field of view (FOV).  We excluded from further analysis time intervals
during which the count rate in the 10-12 keV energy band was higher
than 14.0 \cnts. The resulting effective exposure time was 14.02 ks.
We extracted spectra and relevant light curves in the 0.1-2, 2-10, and
0.1-10 keV energy bands. Since the X-ray count rate of \igrj\ is high,
we corrected the product for pile-up extracting the source in an
annulus with an inner radius of 1.25 pixels. The background extraction
region was chosen at $\sim 9.5$ arcmin from \ngc, far away from any
sources but still in the same CCD of the EPIC-pn camera.  All EPIC
images and spectra were corrected for Out-of-Time (OoT) events.
EPIC-pn spectra were rebinned before fitting so as to have at least
200~counts per energy bin.

\section{Data analysis and results}
\label{sec:res}

\subsection{Source spectrum}

\begin{figure}
\centering
\includegraphics[width=0.5\textwidth]{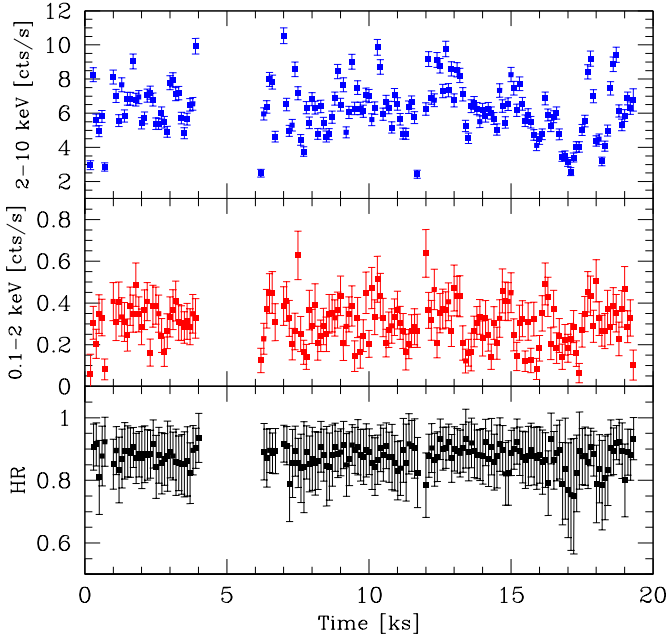}
\caption{\xmm\ Epic-pn background-subtracted light curves of \igrj.
Top and middle panels show light curves extracted in the energy bands
0.1-2 keV (Soft) and 2-10 keV (Hard), respectively. 
The hardness ratio, defined as $\mathrm{HR = (Hard - Soft)/(Hard + Soft)}$, 
is reported in the bottom panel. The time bin is 100 s. 
}
\label{fig:lcxmm}
\end{figure}

\begin{figure}
\centering
\includegraphics[width=0.5\textwidth]{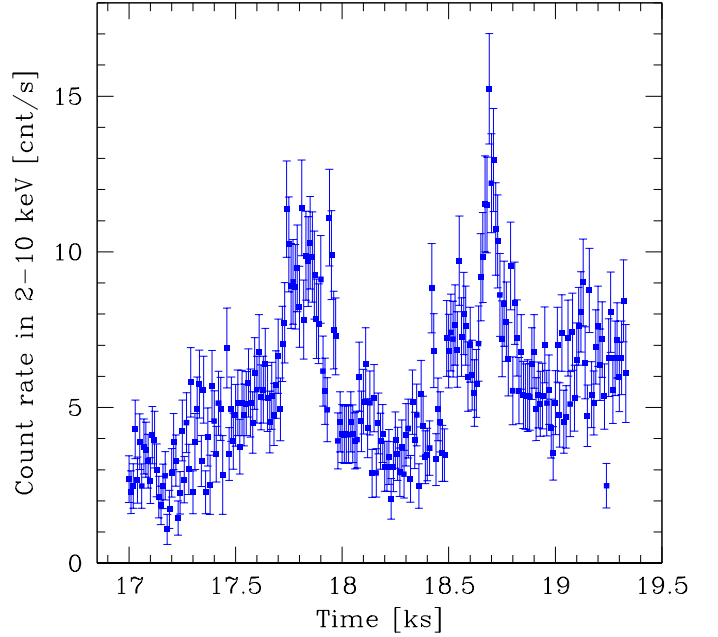}
\caption{Zoom of the \xmm\ Epic-pn background-subtracted light curve
showing the most variable part. The time bin is 10 s. 
}
\label{fig:lczoom}
\end{figure}

We extracted the \xmm\ Epic-pn light curves in the 0.1-2 keV (\soft)
and 2-10 keV (\hard) energy bands with time bins of 10 s.  These two
light curves (rebinned to 100s) and the hardness ratio $HR={\rm (Hard
  - Soft)/(Hard + Soft)}$ are shown in Fig.~\ref{fig:lcxmm}, where a
short time-scale variability is clearly visible. The variability time
scale $\tau$ between two points characterised by the highest
difference in their count rates in 2-10 keV is equal to $90 \pm 5$ s
($\Delta$Ampl$ = 10.5 \pm 2.0$ \cnts, found between 17 and 19.5 ks in
Fig.~\ref{fig:lczoom}).  The count rate in the 2-10 keV band is $6.33$
\cnts, in average, with an observed standard deviation of $2.02$
\cnts. It varies between $1.08 \pm 0.48$ and $15.33 \pm 1.77$ \cnts.


The hardness ratio does not show any significant variability
suggesting that the spectral shape does not vary with flux on short
time scales.  To check this we extracted two spectra by
selecting time intervals in which the count rate in the 2-10 keV
energy band is $< 6.3 $ and $> 6.3$ \cnts. The first spectrum (with
the effective exposure time 5.9~ks) could be fitted by an absorbed
power law. The resulting spectral parameters are $\mathrm{N_H} =
(7.43\pm0.14) \times 10^{22} \mathrm{cm^{-2}}$ and $\Gamma = 2.32\pm
0.04$ for a 2-10 keV X-ray flux of $4.95 \times 10^{-11}$ \ergscm. The
spectrum extracted at a higher count rate (effective exposure time 8.1
ks) could be described using the same model, providing $\mathrm{N_H} =
(7.11\pm0.10) \times 10^{22} \mathrm{cm^{-2}}$ and $\Gamma = 2.13\pm
0.03$ for an 2-10 keV X-ray flux of $6.94 \times 10^{-11}$ \ergscm.

\begin{figure}
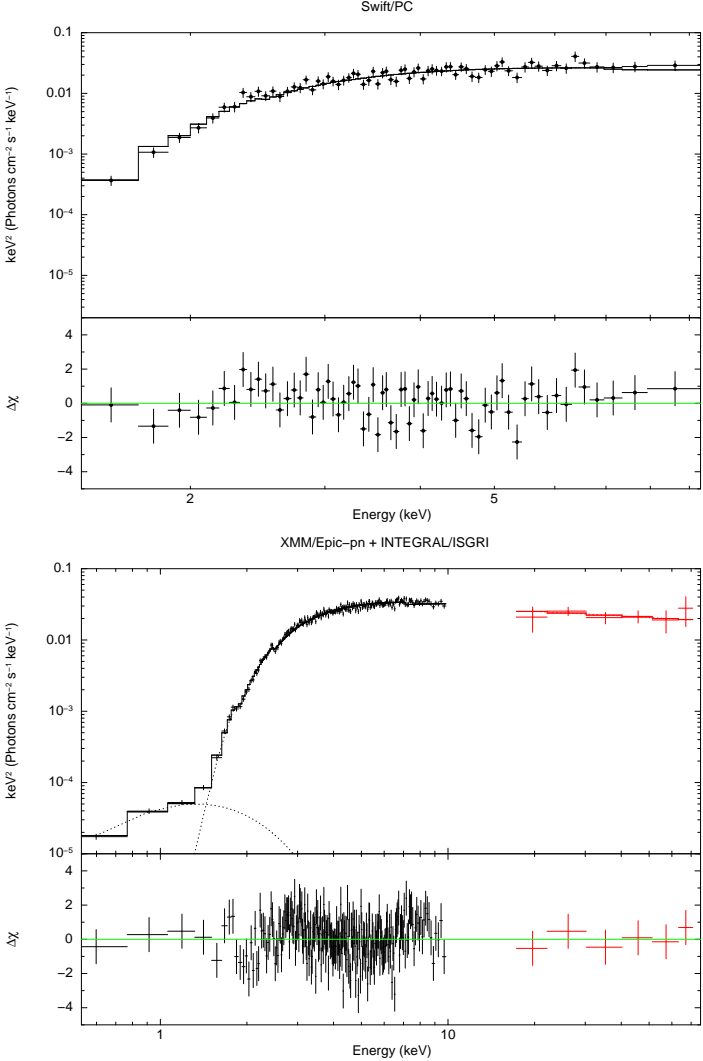

\centering
\includegraphics[height=0.5\textwidth,angle=270]{AA.2012.20664RR_Fig6a.ps}
\includegraphics[height=0.5\textwidth,angle=270]{AA.2012.20664RR_Fig6b.ps}
\caption{
Unfolded spectrum of \igrj\ from \swift\ observations made in
  PC mode (top) and unfolded \xmm/Epic-pn and \integral/ISGRI spectra
  (bottom). The best-fit model (an absorbed power law in all spectra
  plus an unabsorbed black-body emission below 1 keV in case of
  Epic-pn observation) is shown by solid lines in both upper panels.
  Lower panels indicate the residuals ($\Delta \chi$).
}
\label{fig:spec}
\end{figure}

These spectral parameters indicate that the source spectrum did not
change significantly; therefore, we also extracted the average
spectrum of the complete \xmm\ observation (Fig.~\ref{fig:spec},
bottom) that can be represented by the parameters $\mathrm{N_H} =
(7.21\pm0.08) \times 10^{22} \mathrm{cm^{-2}}$ and $\Gamma = 2.19\pm
0.03$. The average 2-10 keV X-ray flux is $6.09 \times 10^{-11}$
\ergscm (see also Table~\ref{tab:fit}).

\swift\ observed \igrj\ almost one week earlier (i.e. on January, 13).
We performed a similar analysis to the one for \xmm\ and found consistent
results (Fig.~\ref{fig:spec}, upper panel). The extracted 1-10 keV
spectrum could be represented by $\mathrm{N_H} = (6.53\pm0.38) \times
10^{22} \mathrm{cm^{-2}}$ and $\Gamma = 2.36\pm 0.14$ for a 
2-10 keV X-ray flux of $4.97 \times 10^{-11}$ \ergscm.

The \isgri\ spectrum of the source in the 17.3-80 keV energy bands was
extracted from observations performed January 2-11 (effective
exposure of 100 ks).  The 17.3-80 keV spectrum could be fitted
($\chi^2$ by d.o.f. = 0.40) by using a power-law model with $\Gamma =
2.17$ for an average hard X-ray flux of $\mathrm{F(17.3-80 keV)}=5.44
\times 10^{-11}$ \ergscm.  A common fit of the ISGRI and \xmm\ data
(with free normalisation) could be obtained, leading to $\mathrm{N_H}
= (7.39 \pm 0.10) \times 10^{22} \mathrm{cm^{-2}}$, $\Gamma = 2.22 \pm
0.03$, and $\mathrm{F(17.3-80 keV)}= (5.408 \pm 0.015) \times
10^{-11}$ \ergscm\ (Fig. \ref{fig:spec}).

\begin{table*}
\caption[]{The best-fit parameters to the \swift\ and \xmm\ observations
in January 2011. The fit to the 1-10 keV energy range.}
\label{tab:fit}
\centering
\begin{tabular}{c l c c c c c} 
\hline
\hline
Satellite/ & Data & \NH & \G & Pl. Norm. at 1 keV  & $F_{2-10 \rm keV}$ & $\chi^2$/d.o.f.\\
$T_{\rm exposure}$ & & ($10^{22}$ cm$^{-2}$) & & 
($10^{-2}$ Photons keV$^{-1}$ cm$^{-2}$ s$^{-1}$) & ($10^{-11}$ erg cm$^{-2}$ s$^{-1}$)\\
\hline
\swift/5.1ks & All data & $6.53 \pm 0.38$& $2.36\pm 0.14$ & $5.69\pm1.33$ & 
            $4.968 \pm 0.130$ & 78.8/81\\
\hline
{\it XMM}/5.9ks & Count rate $< 6.3$   & $7.43\pm0.14$ & $2.32\pm 0.04$ & $5.50\pm0.39$ & 
            $4.945 \pm 0.077$ & 90.9/81\\
{\it XMM}/8.1ks & Count rate $\ge 6.3$ & $7.11\pm0.10$ & $2.13\pm 0.03$ & $5.54\pm0.28$ & 
            $6.937 \pm 0.023$ & 193.7/150\\
{\it XMM}/14ks & All data & $7.21\pm0.08$ & $2.19\pm 0.03$ & $5.45\pm0.23$ & 
            $6.088 \pm 0.020$ & 349.5/225\\
\hline
\end{tabular}
\end{table*}

\subsection{Intrinsic diffuse emission}

The \xmm/Epic-pn spectrum shows an X-ray excess below $\sim 1.2$ keV
(Fig.~\ref{fig:spec}), which can be fitted by a black body absorbed by
the Milky Way leading to a temperature of $T=0.33\pm 0.04$ keV and a
flux of $\mathrm{F_{bbody}(0.01-10 keV)} = 1.12 \times 10^{-13}$
\ergscm.  This black body is not a physical model but does provide a good
representation of the excess. Its flux is in good agreement with the
diffuse emission radiated by galaxies \citep{BG2011,Jia2012}, and not
very far from the \einst\ upper limit and to the residual emission
detected by \swift\ XTE in June 2012 ($\mathrm{F_{bbody}(0.5-2 keV)} =
8 \times 10^{-14}$ \ergscm).

This soft excess is completely unrelated to the nuclear flaring
activity since the nucleus is very strongly absorbed. This emission
comes from a region outside of the dusty torus of \ngc.


\subsection{Duration and frequency of the flare}

Because the source was still detected significantly in July 2011 by
\integral, the flare probably lasted for 150 days at a level above $2
\times 10^{-11}$ \ergscm. \maxi\ observed the source continuously, but
at a lower sensitivity.  Therefore the flare was just detected by
\maxi\ close to the peak in January 2011.

Such a flare of \igrj\ with a duration of 150 days has never been
detected in the past ten years by \integral, and it could have been missed
only between August 2006 and December 2007 when \integral\ did not
observe that region of the sky. The flare of \ngc\ is therefore
exceptional, and no such flare was detected for a continuous period of
at least 1300 days.

\subsection{Black-hole mass}

The mass of the central black hole, \mbh, of \ngc\ can be estimated
using methods based on the X-ray variability time scale.

The upper limit of \mbh\ can first be estimated assuming that the
shortest time variability $\tau$ is related to the innermost stable
circular orbit (ISCO). The \xmm\ light curve observed in 2-10 keV band
indicates $\tau < 90 \pm 5$ s.  Additionally, assuming a non-rotating
black hole, we can write that c$\tau = R_{\rm Schw} = 2G\mmbh/$c$^2$.
The observed variability time scale corresponds to $\mmbh \le
9.6\times 10^6 \mMsun$.

The second technique is based on the X-ray excess variance
measurements.  This method uses the relationship between the
black-hole mass and the X-ray variability $\mmbh = C (T - 2\Delta
t)/\sigma^2_{\rm nxs}$, where $T$ is the duration of the X-ray light
curve and $\Delta t$ is the bin size, both in seconds, and
$\sigma^2_{\rm nxs} = \sum^N_{i=1} [(x_i - \bar{x})^2 -
\sigma^2_{\mathrm{err},i}]/(N \bar{x}^2)$ is the normalised excess
variance \citep{Nandra97,Vaughan2003,ONeill2005}.  We divided the 2-10
keV \xmm\ light curve into two parts with length T$\sim$ 4 ks and
$\sim$10 ks in order to avoid the gap of 5 ks in the light curve.  The
estimated mean normalised excess variance, $\langle \sigma^2_{\rm nxs}
\rangle$, of the 10 s bin light curve is $0.12 ^{+0.09}_{-0.04}$.
Thus, the calculated central black hole mass in \ngc\ is $\mmbh =
2.3^{+1.1}_{-1.0} \times 10^5 \mMsun$, where we assumed $C=1.42$ (i.e.
that the mass of Cyg X-1 is equal to $14.8 \pm 1.0$ $\mMsun$,
\citealt{Orosz2011}). The errors on the mass were estimated 
by performing Monte Carlo simulations \citep[see details in][]{Niki2006}.
The systematic errors, related to the calibration of the excess variance 
versus mass relationship, are larger and the black-hole mass could
range in the interval $10^4-10^6 \mMsun$ \citep{ONeill2005}.

\section{Discussion}
\label{sec:dis}

\subsection{Phenomenology}

We first consider the possibility that the observed variability is
driven by absorption, i.e. that a hole opened on the line of sight in
the AGN absorbing torus. The [O III] line intensity in \ngc\ is
$5.40\times 10^{-15}$ \ergs \citep{HFS97}.  According to the X-ray --
[O III]$_{\lambda5007}$ relationship \citep{Panessa2006} the estimated
unabsorbed 2-10 keV flux should be $\simeq 6 \times 10^{-14}$ \ergs.
However, since the X-ray flux observed in January 2011 was almost 1000
times larger (Table~\ref{tab:fit}), the absorption explanation is
ruled out. A strong increase in the X-ray flux, by a factor $10^3$, is
required.  We also note that the source is not Compton thick,
($N_H\sim 7\times 10^{22}$ cm$^{-2}$, see Table~\ref{tab:fit}),
indicating that it should be bright at hard X-rays, when active.

\begin{figure}
\centering
\includegraphics[width=0.5\textwidth]{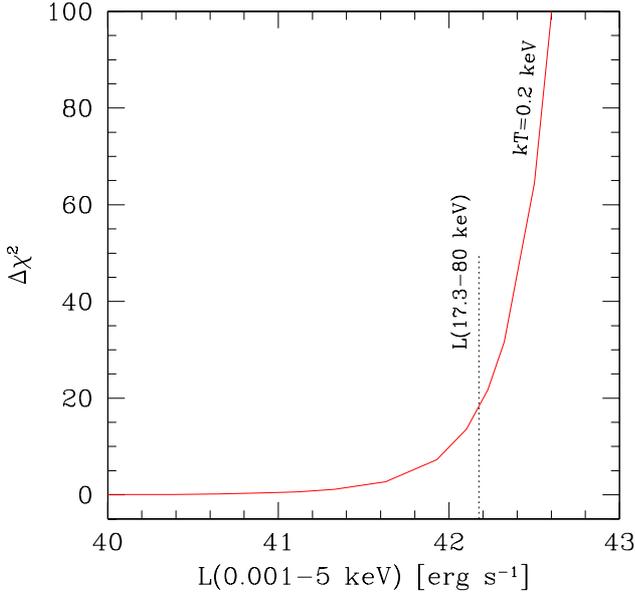}
\caption{
  Increase in $\chi^2$ related to an additional soft X-ray component
  (black-body with temperature of 0.2 keV) increasing in luminosity,
  when fitted by the same model as used in figure~\ref{fig:spec}. A soft
  X-ray component stronger than $5 \times 10^{42}$ erg s$^{-1}$
  should have been detected, if present. }
\label{fig:softX}
\end{figure}

Supernova explosions and expanding winds may emit hard X-ray flares
\citep{IL2003,ChD2009,Saxton2012}. Supernova peak X-ray luminosities
have been observed up to $10^{40}$ erg/s with a decline usually
following a $t^{-(0.5-1)}$ scaling.  The peak X-ray luminosity of
\igrj\ is 100 times brighter, and it declined much faster than
observed in supernovae. This, together with the position of \igrj\ at
the very centre of \ngc, makes the supernova interpretation of the
observed flare very unlikely.

A brightening of the Seyfert nucleus by a factor $10^3$ should therefore 
be explained by a sudden increase in the accretion rate. Tidal disruptions 
of objects by a massive black hole
\citep[see][]{Komossa2004,Burrows2011,Cenko2012,Saxton2012}
have been modelled and simulated by various authors
\citep[e.g.][]{Rees88,EK89,Ulmer99,AlexKum2001,AlexLiv2001,
LNM2002,LKP2009}. 
The induced emission follows a power law decline with a characteristic 
slope of $-5/3$, corresponding to our observations.

The peak of the observed 17.3-80 keV luminosity is $1.5 \times
10^{42}$ \ergs. Most of luminosity of a tidal disruption event is
expected to be released in the soft X-rays (and absorbed by the torus
in \ngc).  We can evaluate the maximum soft X-ray emission allowed by
the data by adding a soft thermal component to the model and determine
for which flux such a component would have been detected.
Figure~\ref{fig:softX} shows the increase in the $\chi^2$ obtained
depending on the soft component luminosity. For a temperature of $kT =
0.2$ keV, which is expected in our case \citep[see eq. 9
in][]{Ulmer99}, the soft component cannot be more than 10 times
brighter than the hard X-ray component. The soft component could of
course be brighter if it would peak at lower energies, which is
however unexpected, especially for a low-mass black hole.

Assuming that the thermal emission is ten times brighter than the
hard X-ray emission, the tidal event luminosity reached the Eddington 
luminosity at maximum with $L_{\rm flare}/L_{\rm Edd}\approx 0.6$, 
as observed in other tidal disruption events \citep{LNM2002}. The 
total energy radiated is then of the order of $10^{50}$ ergs, which 
corresponds to the energy released by the accretion of 0.5 Jupiter 
mass $(\mMjup)$.

\begin{figure}
\centering
\includegraphics[width=0.5\textwidth]{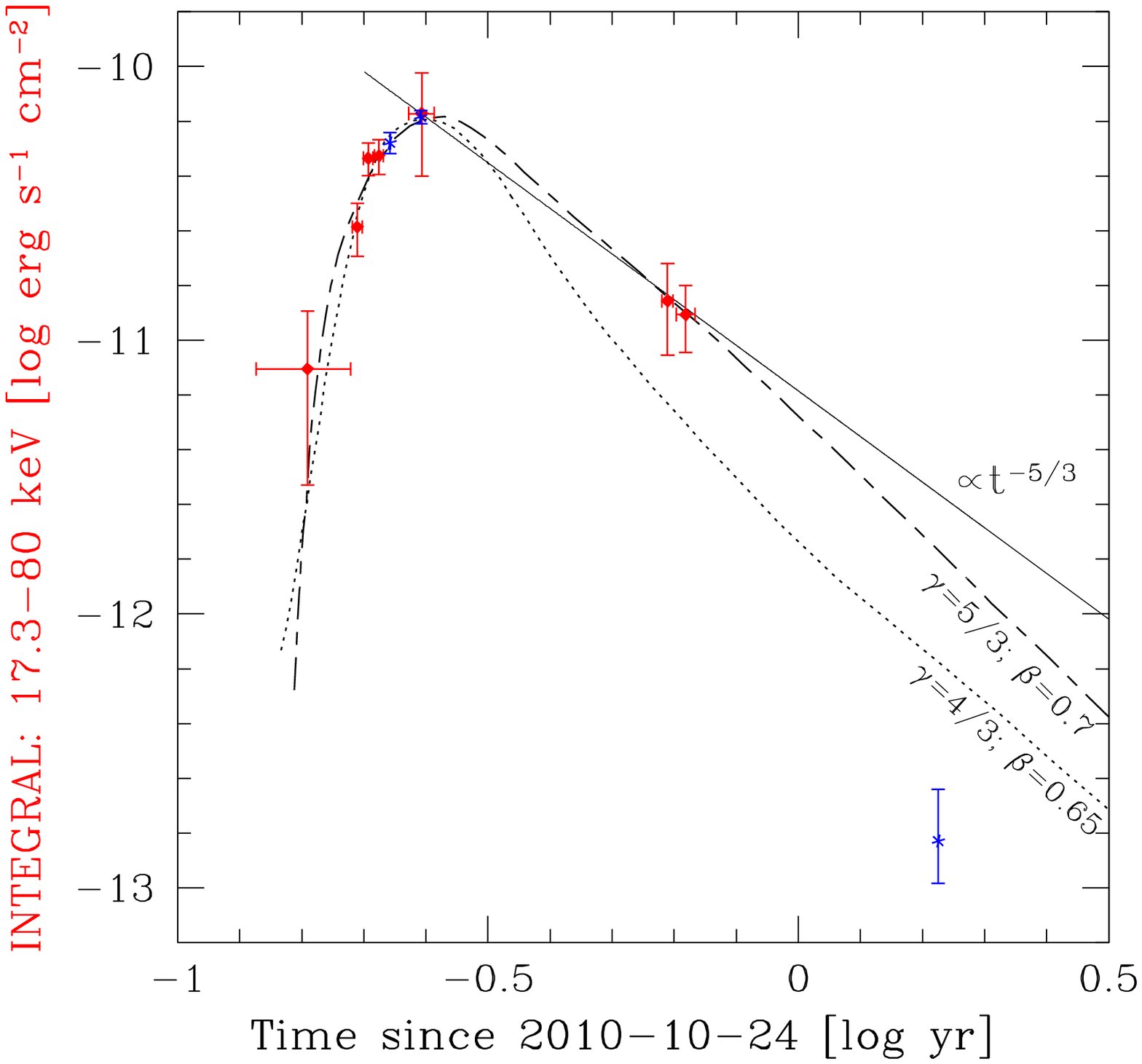}
 \caption{
The light curve of \igrj\ observed in the 17.3-80 keV energy band. 
Red squared points refer to \integral\ data, blue crosses show
\swift\ and \xmm\ observations. The solid line show a model of the form 
$(t - t_D)^{-5/3}$, as expected for fallback of material after a tidal 
disruption event. The long-short dash and dotted lines indicate the
predictions of simulations for the disruption of a sub-stellar object 
($\gamma= 5/3$; $\beta=0.7$) and, respectively of a star 
($\gamma=4/3$; $\beta=0.65$) \citep{GRR2012}.
}
\label{fig:decl}
\end{figure}

We fitted the decline of the \igrj\ X-ray light curve with the
$(t - t_{\rm D})^{-5/3}$ law where $t_{\rm D}$ is the time of the
initial tidal disruption (Fig.~\ref{fig:decl}). The peak luminosity of
the flare occurred on January 22, 2011 according to \maxi\ and
\integral.  It turns out from our fit that the beginning of the
tidal disruption occurred $\simeq$ 60-100 days before the peak of the
X-ray flare, marking the heating of the debris at the vicinity of the
black hole.  A similar delay was mentioned by \citet{LNM2002} as the
time $\Delta t_1$, since the most bounded material returns to the
pericentre.

\subsection{Comparison with tidal disruption simulations}

Detailed hydrodynamic simulations of tidal disruptions were 
performed by \citet{GRR2012} and their results parametrised.
The rate of mass falling on the black hole, $\dot M(t)$, the time of
the peak accretion, $t_{\rm peak}$, as well the decay power-law index,
$\Gamma$, depend on a few parameters: the structure and mass of the
disrupted object and the minimum distance to the black hole.
  
We compared the simulations to our data assuming that the disrupted
object is either a star or a sub-stellar object with polytropic index
$\gamma$ of 4/3 or 5/3, respectively. We used the parametrisation
included into the Appendix of \citet{GRR2012} for different impact
parameters $\beta \equiv r_{\rm T}/r_{\rm P}$, where $r_{\rm T}$, and
$r_{\rm P}$ are the tidal and the pericentric radii, respectively. We
also assumed \mstar -- \rstar\ relations for the disrupted object
valid for sub-stellar objects or stars according to \citet{ChB2000}.

The observational constraints, $t_{\rm peak}\simeq 0.2$ yr and the
peak accretion rate $\dot M_{\rm peak}\simeq$ 2.5 \Mjup/yr (under the
assumption of a hard X-ray radiation efficiency of 10\%), are
sufficient to constrain the mass of the disrupted object.  For a
black-hole mass of $2.3 \times 10^5 \mMsun$, the mass of the disrupted
object turns out to be either 14-16 $\mMjup$ (for $\gamma=5/3$ and
$\beta = 0.6-1.9$) or $10-15\mMsun$ (for $\gamma=4/3$ and $\beta
\simeq 0.65$). For a black-hole mass increasing to $10^6 \mMsun$, the
mass of the disrupted object would be $25-28 \mMjup$ (up to $75\mMjup$
for $10^7 \mMsun$) for a sub-stellar object or $1-3\mMsun$ for a star
($\beta \simeq 0.7$). The $\beta$ parameter is not constrained in the
case of a sub-stellar object, while a very narrow range is required
to disrupt a star. The fraction of the object reaching the black hole
would be very small in the case of a stellar disruption.
 
 
The hydrodynamic simulations also predict the detailed variability of
the accreted mass flow with time. In the case of a star
$(\gamma=4/3)$, less centrally condensed than a sub-stellar object,
the accreted mass decreases much faster after the peak. This is
illustrated in figure~\ref{fig:decl}, where the observed flux behaves
similarly to the predictions for the disruption of a
sub-stellar object and contrasts with those obtained for a star.
  
The hydrodynamic simulations indicate, therefore, that the tidally
disrupted object was probably a $14-30 \mMjup$ sub-stellar object and
that about 10\% of its mass has been accreted on a black-hole
weighting no more than $10^6\mMsun$. We note that a slightly different
equation of state could lead to a mass of the disrupted object as low
as several M$_J$ and that specific simulations of tidal disruption of
sub-stellar objects have not yet been done.

The decline in the hard X-ray flux observed in Fig. \ref{fig:decl}
after 200-500 days can be understood as debris falling back on the
remaining object core and decreasing the emission during the late
evolution \citep[see fig. 8b in][]{GRR2012}. This would indicate that
the disruption was not total. Another explanation could be that the
corona disappears faster than the debris, possibly indicating a change
in the geometry of the accretion flow with time.

%
%

\subsection{Power spectrum}

\begin{figure}
\centering
\includegraphics[width=0.5\textwidth]{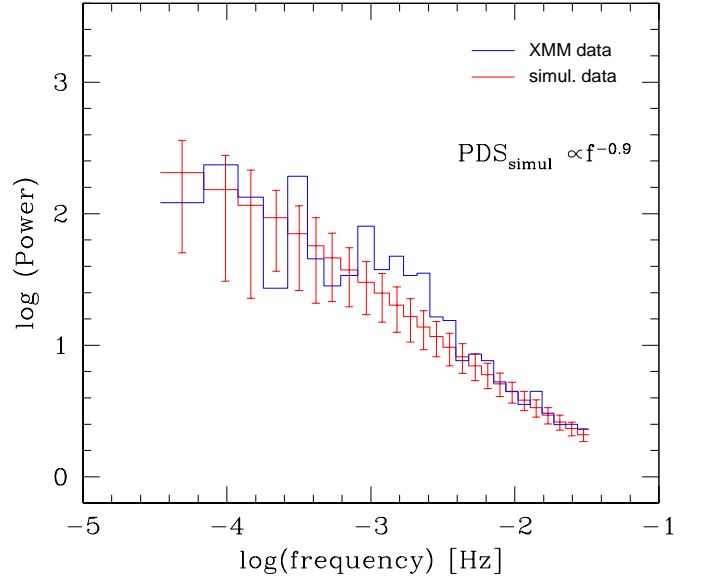}
\caption{The 19.1 ks \xmm\ power spectrum (blue histogram) versus
  simulated 10 000 powers (red histogram with error bars).  The power
  spectrum of the simulated light curves are power law shaped with a
  slope of -0.9. 1$\sigma$ error bars are shown. An excess in the
  observed power near the log(frequency)$=$ -3 may suggest a presence
  of the QPO feature. }
\label{fig:f09}
\end{figure}
\begin{figure}
\centering
\includegraphics[width=0.5\textwidth]{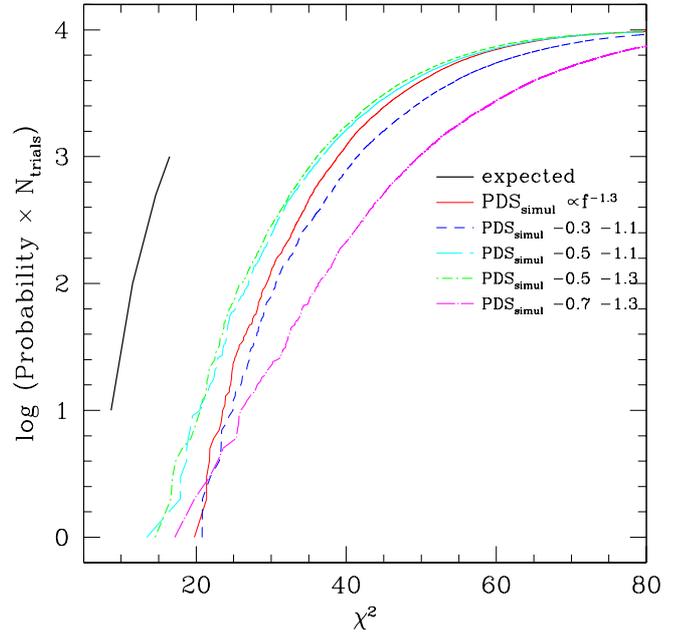}
\caption{Probability multiplied by number of trials ($N_{\rm trials} =
  10000$) against $\chi^2$ for simulated single and broken power-law
  models. The solid black curve points out expected probability of a
  perfect model. The best fit by a single power law is obtained with a
  slope $=$ -1.3. Broken power laws with the low-frequency slope =
  -0.5, the high-frequency slope between -1.1 and -1,3, and with
  broken frequency = 0.001 Hz provide a slightly better description of
  the observed power spectrum.  }
\label{fig:probab}
\end{figure}

The power spectrum of the X-ray light curve obtained by \xmm\ at
the flare maximum (see Fig.~\ref{fig:f09}) shows an excess in the
range 0.0008-0.004 Hz. This excess could be similar to a
quasi-periodic oscillation (QPO) in the accretion flow. The frequency
range of the excess corresponds to the innermost stable circular orbit
for a black hole of a few $\times 10^5 \mMsun$ \citep{ReMc2006}.

To investigate the significance of this excess, we performed
Monte Carlo simulations. We assumed various shapes of the power
spectrum of the source and generated 10'000 light curves for each of
them \citep{TK95}. The spread of the power spectra of all simulated
light curves were then compared to the shape of the observed power
spectrum, as outlined in \citet{UMP2002}.

We tried single power-law power spectra with slopes between -0.7
and -1.5 and broken power-law models with break frequency frozen to
$10^{-3}$ Hz, which differ by the low- and high-frequency slopes.
Figure~\ref{fig:probab} shows the $\chi^2$ distributions obtained for
the best models, together with the expected distribution. The deviation
between the simulated and expected distribution indicates that the
excess can be obtained by chance at a probability not greater than 5\%.
We conclude that a longer \textit{XMM} observation would have been
very useful and note that an instrument like the Large Observatory For
X-ray Timing (LOFT) would have detected several million X-ray photons
during one ISCO period for this tidal event, which would have allowed
a probe of the geometry of the debris falling towards the black-hole
horizon.

\subsection{Tidal events frequency}

The density of stars in the centre of a well developed galaxy is
around $10^2 \mMsun/$pc$^3$ in a region of 100 pc in radius and is
comparable to what is found in globular cluster \citep{MarxPfau92}.
These dense regions allow stellar encounters to create new
gravitationally bound systems or ejections \citep{DavBenz95}. Stars in
the vicinity of a supermassive black hole may escape to infinity after
encounter because their orbits are deflected \citep{AlexKum2001,
  AlexLiv2001}.  Other processes may also play important roles
\citep[see brief review by][]{Alexander2012}. The tidal event rate in
bulges are estimated to be in the range $10^{-6} - 10^{-4}$ yr$^{-1}$
galaxy $^{-1}$ growing to $10^{-2}$ in case of chaotic stellar motions
in a non- axisymmetric gravitational potential \citep{MP2004}. A
reasonable rate for the stellar tidal disruption events is $\sim
10^{-5}$ yr$^{-1}$ galaxy $^{-1}$
\citep{Donley2002,WM2004,Gezari2009}.

The discovery of a population of unbound Jupiter-mass objects has
recently been mentioned by \citet{Sumi2011} and \citet{Delorme2012}.
The number of free-floating massive planets, which are kicked away
from young star-forming regions \citep[e.g.][]{Quanz2010}, has been
estimated \citep{Sumi2011} as one to three times higher than the
number of main-sequence stars. Because of the different equations of
state, the tidal radii are similar for objects of $1 \mMsun$ and $1
\mMjup$ around a massive black hole. Tidal disruptions of planets
could therefore be as frequent as star disruption. Planets could
increase the rate of tidal events in galaxies a few times, but
will generate lower luminosity events on average.

We estimate that \integral\ could detect tidal disruption flares with 
fluxes ten times lower than observed in \igrj\, i.e. from galaxies up to a 
distance of 50 Mpc, for similarly faint events. The number of galaxies 
within D $<$ 50 Mpc is $\sim 5000$. The event rate detectable at hard 
X-rays (by \integral\ or \swift) could therefore be expected to be 
a few events every ten years. The \integral\ and \swift\ 
archives can be used to test these predictions.

%
\begin{acknowledgements}
  Based on observations with INTEGRAL, an ESA project with instruments
  and science data centre funded by ESA member states (especially the
  PI countries: Denmark, France, Germany, Italy, Switzerland, Spain),
  and Poland and with the participation of Russia and the USA.  We thank
  Enrico Bozzo, Laetitia Gibaud, and Claudio Ricci for help with
  \swift\ and \xmm\ analyses; Piotr \.Zycki for giving his software
  template and to Bo\.zena Czerny for discussion about light-curve
  simulations.  MN also thanks the Scientific Exchange Programme
  (Sciex) NMS$^{\rm ch}$ for opportunity of working at ISDC.  This
  research has been supported in part by the Polish NCN grants N~N203
  581240 and 2012/04/M/ST9/00780.
\end{acknowledgements}
\bibliographystyle{aa}
\bibliography{AA.2012.20664RR}

%

\end{document}